\documentclass{jpp}

\usepackage{graphicx}
\usepackage{natbib}
\usepackage{bm}
\usepackage{epsfig}
\usepackage{amsmath}
\usepackage{amsfonts}
\usepackage{amssymb}

\newcommand\etal{\mbox{\textit{et al.}}}

\newsavebox{\astrutbox}
\sbox{\astrutbox}{\rule[-5pt]{0pt}{20pt}}

\title[]{Colliding Alfv\'enic wave packets in MHD, Hall and kinetic simulations}

\author[O. Pezzi et al.]{O. Pezzi$^1$\thanks{Email address for correspondence: oreste.pezzi@fis.unical.it}, T.N. Parashar$^2$, S. 
Servidio$^1$, F. Valentini$^1$, C.L. V\'asconez$^3$, Y. Yang$^2$, F. Malara$^1$, W.H. Matthaeus$^2$ and P. Veltri$^1$}

\affiliation{$^1$Dipartimento di Fisica, Universit\`a della Calabria, 87036 Rende (CS), Italy. \\
$^2$Department of Physics and Astronomy, University of Delaware, DE 19716, USA. \\
$^3$Departamento de F\'isica, Escuela Polit\'ecnica Naciocal, Quito, Ecuador.
}

\pubyear{}
\volume{}
\pagerange{}
\date{?; revised ?; accepted ?. - To be entered by editorial office}

\begin{document}
\maketitle

\begin{abstract}
The analysis of the Parker--Moffatt problem, recently revisited in \citet{pezzi16}, is here extended by including the Hall 
magnetohydrodynamics and two hybrid kinetic Vlasov-Maxwell numerical models. The presence of dispersive and kinetic 
features is studied in detail and a comparison between the two kinetic codes is also reported. Focus on the presence of 
non-Maxwellian signatures shows that - during the collision - regions characterized by strong temperature anisotropy are recovered 
and the proton distribution function displays a beam along the direction of the magnetic field, similar to some recent 
observations of the solar wind.
\end{abstract}

\begin{PACS}
Authors should not enter PACS codes directly on the manuscript, as these must be chosen during the online submission process and 
will then be added during the typesetting process (see http://www.aip.org/pacs/ for the full list of PACS codes)
\end{PACS}

\section{Introduction}
\label{sec:intro}

The interaction of two oppositely propagating Alfv\'enic wave packets has been studied for more than half a century. This 
interaction has been proposed as an elementary step in the analysis of magnetohydrodynamics (MHD) turbulence 
\citep{elsasser50,iroshnikov64,kraichnan65,DMV80a,DMV80b,velli89,sridhar94,goldreich95,NgBhattacharjee,MatthaeusEA99,galtier00,
verdini09,howes13,nielson13}. Indeed, in the framework of ideal incompressible magnetohydrodynamics (MHD), large amplitude 
perturbations in which the magnetic ${\bf b}$ and bulk velocity ${\bf u}$ fluctuations are either perfectly correlated, or 
perfectly anti-correlated, are solutions of the governing equations. To induce nonlinear couplings among the fluctuations, and 
therefore to excite turbulence, it is necessary to simultaneously consider magnetic fluctuations ${\bf b}$ and velocity 
fluctuations ${\bf u}$ that have an arbitrary sense of correlation. This may be accomplished by superposing the two senses of 
correlation, in Alfv\'en units, ${\bf u} = + {\bf b}$ and ${\bf u} = - {\bf b}$.

Based on these considerations, \citet{moffatt78} and \citet{parker79} analyzed the collision of large-amplitude Alfv\'en wave 
packets in the framework of the incompressible MHD, observing that their interaction is limited to the time interval in which they 
overlap. During this temporal window wave packets can transfer energy and modify their spatial structure; however, after the 
collision, packets return to undisturbed propagation without further interactions. The Parker--Moffatt problem has been 
recently revisited in \citep{pezzi16} (hereafter, Paper I) with the motivation to extend its description to the realm of the 
kinetic plasmas. In fact, the scenario described by Parker \& Moffatt is potentially applicable to astrophysical plasmas such as 
solar wind \citep{BelcherDavis71,BrunoEA,verdini09} or solar corona \citep{MatthaeusEA99-coronal,tomc07}, where Alfv\'enic 
perturbations represent one of the main components of fluctuations. However, since such systems often exhibit compressive 
activity as well as dispersion and kinetic signatures \citep{AlexandrovaEA08, bruno13, SahraouiEA07, GaryEA10, marsch06, 
valentini11, servidio12, valentini14, servidio15, he15, roberts16, lion16, perrone16}, it is of considerable interest to include 
these features in the analysis of the {\it Parker} \& {\it Moffatt} problem.

In particular, in Paper I, it has been found that during the wave packets interaction, as prescribed by {\it Parker} \& {\it 
Moffatt}, nonlinear coupling processes cause the magnetic energy spectra to evolve towards isotropy, while energy transfers 
towards smaller spatial scales. Moreover, the new ingredients introduced with the kinetic simulation (Hall and kinetic effects) 
play a significant role and several features of the evolution in the Vlasov case differ with respect to the MHD evolution. 
Here we extend that study to discern the role of dispersive and genuinely kinetic effects, supplementing the previously 
considered MHD and Vlasov simulations, by introducing also an Hall MHD simulation. Moreover, we also examine this basic problem 
by means of a hybrid Particle-in-Cell simulation (HPIC), which allows comparison of two different numerical approaches (HVM and 
HPIC), which refer to the same physical model. We may anticipate that, in the HPIC case, the system dynamics at small scales is 
affected by the presence of particles thermal noise and only the features related to large spatial scales are properly recovered 
during the evolution of the two wave packets. Based on this consideration, we employ mainly the HVM simulation to highlight the 
presence of kinetic effects during the wave packets interaction. In particular, during the collision of the wave packets, the 
proton velocity distribution function (VDF) exhibits a beam along the background magnetic field direction, similar to some solar 
wind observations \citep{he15}. We note that the present paper compares results from four different models in the context of a 
single physical problem, and is therefore also a contribution in the spirit of the ``Turbulence Dissipation Challenge'' that has 
been recently discussed in the space plasma community \cite{ParasharJPP15}.

The paper is organized as follows: in Sec. \ref{sec:models} the theoretical model and the numerical codes are presented. In Sec. 
\ref{sec:numresults}, we compare the several simulations by focusing on the description of some fluid-like diagnostics. Sec. 
\ref{sec:kin}, examines kinetic signatures in the HVM simulation. Finally we conclude in Sec. \ref{sec:concl} by summarizing our 
results.

\section{Theoretical models and numerical approaches}
\label{sec:models}

As discussed above, here we approach the problem concerning the interaction of two Alfv\'enic wave packets by means of fluid and 
hybrid numerical simulations. For problems such as this, the system dimensionality is fundamental: in fact, a proper description 
should consider a three-dimensional physical space (i.e. three-dimensional wave vectors), where both parallel and perpendicular 
cascades are taken into account \citep{howesJPP15,parasharEA15,ParasharEA16}. However, dynamical range of the spatial scales (wave 
numbers) represented in the model is equally important to capture nonlinear couplings during the wave packet interaction. 
Furthermore, performing a kinetic HVM simulation which contemporaneously includes a full $3D$--$3V$ phase space and while 
also retaining a good spatial resolution is too demanding for the present High Performance Computing capability. Given that 
numerous runs are required to complete a study such as the present one, a fully 3D approach would be prohibitive. Therefore we 
adopt a $2.5D$ physical space, where vectorial fields are three-dimensional but their variations depend only on two spatial 
coordinates ($x$ and $y$). It is worth noting that $2.5D$ captures the qualitative nature of many processes very well even though 
there might be some quantitative differences for some processes \citep{KarimabadiSSR2013, WanPRL2015, LiApJL2016}.

The fluid models here considered are MHD and Hall MHD, whose dimensionless equations are:
\begin{eqnarray}
& \partial_t \rho +\nabla \cdot (\rho {\bf u})=0 \label{eq:HMHD1} \\ 
& \partial_t {\bf u} +({\bf u}\cdot \nabla){\bf u}= -\frac{{\tilde \beta}}{2\rho}\nabla (\rho T)+ \frac{1}{\rho}
\left[(\nabla \times {\bf B})\times {\bf B}\right]\label{eq:HMHD2} \\
& \partial_t {\bf B} = \nabla \times \left[ {\bf u}\times {\bf B} -\frac{{\tilde\epsilon}}{\rho} (\nabla \times {\bf B})\times 
{\bf B} \right] \label{eq:HMHD3} \\
& \partial_t T + ({\bf u}\cdot \nabla)T + (\gamma -1)T(\nabla \cdot {\bf u})=0 \label{eq:HMHD4}
\end{eqnarray}
In Eqs. (\ref{eq:HMHD1})--(\ref{eq:HMHD4}) spatial coordinates ${\bf x}=(x,y)$ and time $t$ are respectively normalized to
$\tilde{L}$ and $\tilde{t}_A=\tilde{L}/\tilde{c}_A$. The magnetic field ${\bf B}={\bf B}_0 + {\bf b}$ is scaled to the typical
magnetic field ${\tilde B}$, while mass density $\rho$, fluid velocity ${\bf u}$, temperature $T$ and pressure $p=\rho T$ are
scaled to typical values ${\tilde \rho}$, ${\tilde c}_A={\tilde B}/(4\pi {\tilde \rho})^{1/2}$, ${\tilde T}$ and ${\tilde
p}=2\kappa_B {\tilde \rho}{\tilde T}/m_p$ (being $\kappa_B$ the Boltzmann constant and $m_p$ the proton mass), respectively.
Moreover, ${\tilde \beta}={\tilde p}/({\tilde B}^2/8\pi)$ is a typical value for the kinetic to magnetic pressure ratio; $\gamma
=5/3$ is the adiabatic index and ${\tilde \epsilon} = {\tilde d}_p/{\tilde L}$ (being ${\tilde d}_p={\tilde c}_A/{\tilde 
\Omega}_{cp}$ the proton skin depth) is the Hall parameter, which is set to zero in the pure MHD case. Details about the 
numerical algorithm can be found in \citet{vasconez15,pucci16}.

On the other hand, hybrid Vlasov-Maxwell simulations have been performed by using two different numerical codes: an Eulerian 
hybrid Vlasov-Maxwell code (HVM) \citep{valentini07} and a hybrid Particle-in-cell (HPIC) code \citep{ParasharPP09}. For both 
cases protons are described by a kinetic equation,while electrons are a Maxwellian, isothermal fluid. In the Vlasov model, an 
Eulerian representation of the Vlasov equation for protons is numerically integrated. In PIC method, the distribution function is 
Monte-Carlo discretized and the Newton-Lorentz equations are updated for the ``macro-particles''. Electromagnetic fields, charge 
density and current density are computed on a grid \citep{Birdsall&LangdonBook,DawsonRMP1983}.

Dimensionless HVM equations are:
\begin{eqnarray}
&\partial_t f + {\bf v} \cdot \nabla f + \frac{1}{{\tilde \epsilon}} &\left( {\bf E}+{\bf v}\times{\bf B}\right)\cdot 
\nabla_{\bf v} f =0 \label{eq:hvm1} \\ 
&{\bf E} -\frac{m_e {\tilde \epsilon}^2}{m_p} \Delta {\bf E}  = &-{\bf u_e}\times {\bf B} -\frac{{\tilde \epsilon}{\tilde 
\beta}}{2n} \left( \nabla P_e - \frac{m_e}{m_p}\nabla\cdot {\bf \Pi}\right) +  \nonumber \\
&&+  \frac{m_e}{m_p} \left[{\bf u}\times {\bf B} + \frac{{\tilde \epsilon}}{n} \nabla\cdot \left(n\left({\bf u}{\bf u}-{\bf 
u_e}{\bf u_e}\right)\right) \right]\label{eq:hvm2}  \\
&\frac{\partial {\bf B}}{\partial t}=-\nabla \times {\bf E} \;\;\; ; \;\;\; &\nabla \times {\bf B}={\bf j} \label{eq:hvm3}
\end{eqnarray}
where $f = f({\bf x},{\bf v},t)$ is the proton distribution function. In Eqs. (\ref{eq:hvm1})--(\ref{eq:hvm3}), velocities ${\bf 
v}$ are scaled to the Alfv\'en speed ${\tilde c_A}$, while the proton number density $n=\int f\, d^3v$, the proton bulk 
velocity ${\bf u}=n^{-1}\int {\bf v} f\, d^3v$ and the proton pressure tensor ${\Pi}_{ij}=n^{-1}\int (v-u)_i\,(v-u)_j f\, 
d^3v$, obtained as moments of the distribution function, are normalized to $\tilde{n}=\tilde{\rho}/m_p$, $\tilde{c}_A$ and 
${\tilde p}$, respectively. The electric field $\bf{E}$, the current density ${\bf j}= {\bf \nabla}\times{\bf B}$ and the 
electron pressure $P_e$ are scaled to $\tilde{E}=(\tilde{c}_A \tilde{B})/c$, $\tilde{j}=c\tilde{B}/(4\pi\tilde{L})$ and 
$\tilde{p}$, respectively. Electron inertia effects have been considered in Ohm's law to prevent numerical instabilities (being 
$m_e/m_p=0.01$, where $m_e$ is the electron mass), while no external resistivity $\eta$ is introduced. A detailed description of 
the HVM algorithm can be found in \citet{valentini07}. On the other hand, the hybrid PIC run has been performed using the P3D 
hybrid code \citep{ZeilerJGR02} and all the numerical and physical parameters are the same as the HVM run. The code has been 
extensively used for reconnection and turbulence \citep[e.g.][]{ParasharPP09, MalakitGRL09}. 

\begin{figure*}
\epsfxsize=\textwidth \centerline{\epsffile{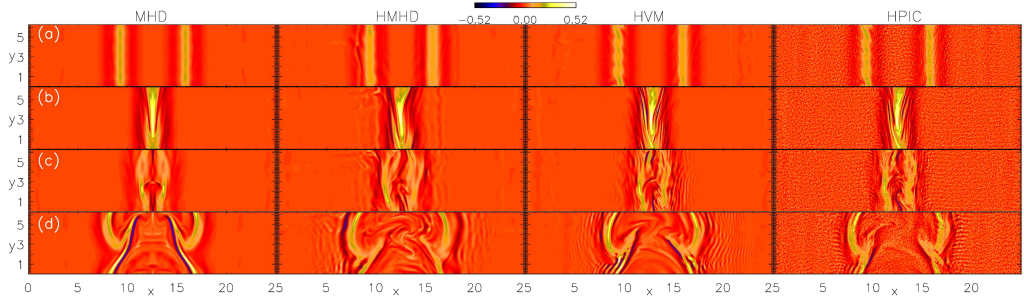}}   
\caption{(Color online) Contour plot of the out of plane component of the current density $j_z(x,y)$ at several time instant 
$t=29.5$ (a), $t=\tau=58.9$ (b), $t=70.7$ (c) and $t=98.2$ (d). From left to right, each column refers to the MHD, HMHD, HVM and 
HPIC cases, respectively. For the HPIC simulation, $j_z(x,y)$ has been smoothed in order to remove particle noise.}
\label{fig:jz}
\end{figure*}

In both classes of performed simulation (fluid and kinetic), the spatial domain $D(x, y) = [0, 8\pi] \times [0, 2\pi]$ is 
discretized with $(N_x,N_y) = (1024, 256)$ in such a way that $\Delta x=\Delta y$ and spatial boundary conditions are periodic. 
For the HVM run, the velocity space is discretized with an uniform grid with $51$ points in each direction, in the region $v_i = 
[-v_{max}, v_{max}]$ (being $v_{max} = 2.5 {\tilde c_A}$) and velocity domain boundary conditions assume $f=0$ for $|v_i|>v_{max}$ 
($i=x,y,z$); while, in the HPIC case, the number of particles per cell is $400$. Moreover $\beta_p = 2v_{th,p}^2/{\tilde c_A}^2= 
{\tilde \beta}/2 =0.5$ (i.e. $v_{max} = 5 v_{th,p}$), ${\bf u_e} = {\bf u} - {\tilde \epsilon} {\bf j}/n$, ${\tilde 
\epsilon}=9.8\times 10^{-2}$, $k_{d_p} = {\tilde \epsilon}^{-1} \simeq 10$ and $k_{d_e} = \sqrt{m_p /m_e}\times {\tilde 
\epsilon}^{-1} \simeq 100$. The background magnetic field is mainly perpendicular to the $x-y$ plane: ${\mathbf B_0} = B_0 (\sin 
\theta, 0, \cos \theta)$, where $\theta=\cos^{-1} \left[\left(\mathbf{B_0} \cdot {\hat{\bf z}} \right)/B_0\right] = 6^\circ$ and 
$B_0=|\mathbf{B_0}|$. 

In the initial conditions, ions are isotropic and homogeneous (Maxwellian velocity distribution function in each spatial point) 
for both kinetic simulations. 

Large amplitude magnetic ${\bf b}$ and bulk velocity ${\bf u}$ perturbations are introduced, while no density perturbations 
are taken into account (which implies nonzero total pressure fluctuations). Initial perturbations consist of two Alfv\'enic wave 
packets with opposite velocity-magnetic field correlation. The packets are separated along $x$ and, since $B_{0,x}\neq 0$, 
they counter-propagate. The nominal time for the collision, evaluated with respect to the center of each wave packet, is 
$\tau\simeq58.9$. 

The magnetic field perturbation $\mathbf{b}$ has been built in such a way that $\mathbf{B_0} \cdot \mathbf{b} = 0$ is satisfied in 
each spatial point. Then the velocity field perturbation $\mathbf{u}$ is built by imposing that $\mathbf{u}$ and $\mathbf{b}$ are 
correlated (anti-correlated) for the wave packet which moves against (along) $B_{0x}$. A detailed discussion about the properties 
of the initial perturbations can be found in Paper I. The condition $B=|{\bf B}|= const$ is not satisfied by our initial 
perturbations, while this condition would be a requirement in defining a large amplitude Alfv\'en mode in the context of a 
compressible MHD model. This suggests that pressure and density fluctuations are generated during the wave packet evolution.  

\begin{figure}
\epsfxsize=11cm \centerline{\epsffile{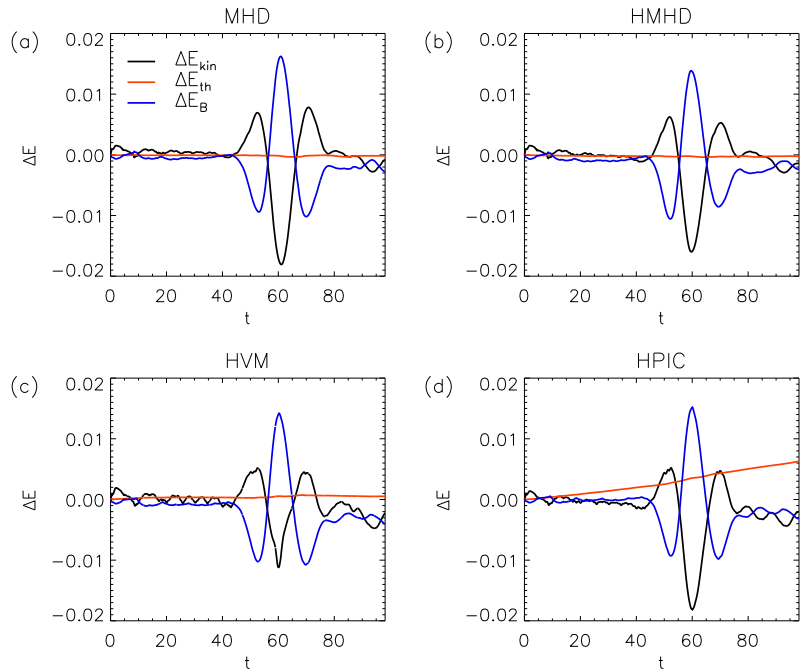}}   
\caption{(Color online) Temporal evolution of the energy terms: $\Delta E_{kin}$ (black), $\Delta E_{th}$ (red) and $\Delta 
E_{B}$ (blue) for the MHD, HMHD, HVM and HPIC runs. }
\label{fig:conserv}
\end{figure}

The perturbations intensity is $\langle b\rangle_{rms}/B_0 =0.2$, therefore the Mach number is $M_s = \langle u\rangle_{rms}/ 
v_{th,p} = 0.4$. The intensity of fluctuations with respect to the in-plane field $B_{0x}$ is quite strong, with a value of about 
$2$. This last parameter can be associated with $\tau_{nl}/ \tau_{coll}$ (characteristic nonlinear time $\tau_{nl}$; 
characteristic collision time $\tau_{coll}$), whose value gives insight about the type of turbulence which could be generated. 
Here $\tau_{nl}/ \tau_{coll}\simeq 0.5$, hence nonlinear effects are important to approach a strong turbulence scenario. 

\section{Numerical results: a comparison between several codes}
\label{sec:numresults}

In this Section we focus on the description of the results of the four different simulations (MHD, HMHD, HVM and HPIC) by 
focusing on some ``fluid''-like diagnostics which help to understand the system dynamics and, also, to compare the numerical 
codes.

Figure \ref{fig:jz} reports a direct comparison between the simulations, showing the contour plots of the out-of-plane component 
of the current density $j_z=({\bf \nabla}\times{\bf B})\cdot {\hat{\bf z}}$. Vertical columns from left to right in Fig. 
\ref{fig:jz} refer to MHD, HMHD, HVM and HPIC simulations, respectively; while each horizontal row refers to a different time 
instant: $t=29.5$ (a), $t=\tau=58.9$ (b), $t=70.7$ (c) and $t=98.2$ (d).  

Significant differences are recovered in the MHD case with respect to the HMHD, HVM and HPIC runs. While the MHD evolution 
is symmetric with respect to the center of the $x$ direction, in the other cases this symmetry is broken also before the wave 
packets interaction due to the presence of dispersive effects which differentiate the propagation along and across the 
background magnetic field. Moreover, during the wave packets overlap [Fig. \ref{fig:jz}(b)], smaller scales structures are formed 
in the HMHD and the HVM cases with respect to the pure MHD evolution, while the HPIC run - despite it recovers several 
significant features of the wave packets interaction - suffers the presence of particles thermal noise, which has been 
artificially smoothed out in Fig. \ref{fig:jz}.

After the collision [Fig. \ref{fig:jz} (c) and (d)], the difference between the MHD and the other simulations becomes stronger.  
In particular, some vortical structures at the center of the spatial domain are recovered in the HMHD and HVM cases, in contrast 
to the pure MHD case. Moreover, the Vlasov simulation exhibits some secondary ripples in front of each wave packet whose nature 
could be related to some wave-like fluctuations. These secondary, low-amplitude ripples are not recovered in the other 
simulations: in fact, they cannot be appreciated in the HPIC run where the noise prevents the formation of such structures while, 
in the Hall simulation, they are only roughly visible. The nature of these low-amplitude ripples is compatible with a KAW-like 
activity and will be reported in detail in a separate paper.

In order to compare models and codes, we display, in Fig. \ref{fig:conserv}, the temporal evolution of the energy variations 
$\Delta E$. Black, red and blue lines indicate respectively the kinetic $\Delta E_{kin}$, thermal $\Delta E_{th}$ and magnetic 
$\Delta E_B$ energy variations, while each panel from (a) to (d) refers to the MHD, HMHD, HVM and HPIC runs, respectively. The 
evolution of $\Delta E_{kin}$ and $\Delta E_B$ is quite comparable in all the performed simulations and, in the temporal range 
where the wave packets collide, magnetic and kinetic energy is exchanged. On the other hand, the evolution of the thermal 
energy $\Delta E_{th}$ differs in the HPIC case compared to the other simulations. Indeed, $\Delta E_{th}$ remains quite close to 
zero for all the simulations except for the HPIC run, where it grows almost linearly for the presence of numerical noise. It is 
worth to note that, as the number of particles increases, the evolution of $\Delta E_{th}$ would get closer to the one obtained in 
the MHD, HMHD and HVM simulations.

\begin{figure*}
\epsfxsize=\textwidth \centerline{\epsffile{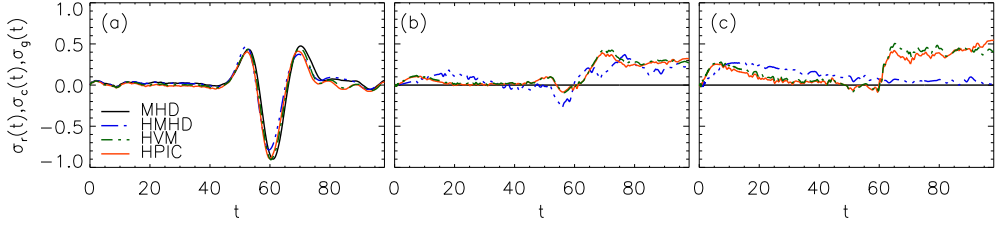}}   
\caption{(Color online) Temporal evolution of the normalized residual energy $\sigma_r(t)$ (a), cross helicity $\sigma_c(t)$ (b) 
and generalized cross helicity $\sigma_g(t)$ (c). In each panel black, blue, green and red lines indicate the MHD, HMHD, HVM and 
HPIC simulations, respectively.}
\label{fig:hchg}
\end{figure*}

The scenario described by Moffatt and Parker is also based on the property,  in ideal incompressible MHD, that two wave packets 
separately conserve energy, which is equivalent to conservation of both total energy and cross helicity $\sigma_c$. It is natural 
therefore to examine evolution of cross helicity, as well as the evolution of the residual energy $\sigma_r$, which gives 
information about the relative strength of magnetic fluctuations and the fluid velocity fluctuations. Figures \ref{fig:hchg}(a--b) 
show the temporal evolution of normalized residual energy $\sigma_r$ (a) and the normalized cross-helicity $\sigma_c$ (b). These 
quantities are defined as follows: $\sigma_r= (e^u - e^b) / (e^u + e^b) = 2 e^u/(e^++e^-)$, where $e^r=e^u-e^b$, $e^\pm=\langle 
(\mathbf{z^\pm})^2\rangle/2$ ($\mathbf{z^\pm}=\mathbf{u}\pm\mathbf{b}$), $e^u=\langle\mathbf{u}^2\rangle/2$ and 
$e^b=\langle\mathbf{b}^2\rangle/2$; $\sigma_c= (e^+ - e^-) / (e^+ + e^-) = 2 e^c /(e^u+e^b)$, being $e^c=\langle 
\mathbf{u}\cdot\mathbf{b}\rangle/2$. In each panel of Fig. \ref{fig:hchg}, black, dashed blue, dashed green and red lines refer to 
MHD, HMHD, HVM and HPIC cases, respectively. 

Figure \ref{fig:hchg} (a) shows the evolution of the normalized residual energy $\sigma_r$, which is similar in all the 
simulations. In particular $\sigma_r\simeq 0$ in the initial stage,it oscillates during the wave packets overlapping,
and finally it returns to $\sigma_r\simeq 0$ after the collisions. The $\sigma_r$ oscillations are well correlated with the 
oscillations of $\Delta E_{B}$ and $\Delta E_{kin}$ seen in Fig. \ref{fig:conserv}.

Deeper insights are revealed by the evolution of the cross-helicity $\sigma_c$. Indeed, for ideal incompressible MHD, the cross 
helicity remains constant, and for this initial condition, $\sigma_c = 0$. Here, $\sigma_c$ is well-preserved in the MHD run 
despite this simulation being compressible. This means that the compressible effects, introduced here by the fact that initial 
perturbations are not pressured balanced, are not strong enough to break the $\sigma_c$ invariance. On the other hand, for the 
remaining simulations (HMHD, HVM and HPIC), $\sigma_c$ is not preserved: i) it shows a jump around $t=\tau=58.9$, due to the 
presence of kinetic and dispersive effects, and ii) there is an initial growth of $\sigma_c$ followed by a relaxation phase. 
It seems also significant to point out that, the initial growth of $\sigma_c$ occurs faster in the kinetic cases compared to the 
HMHD one. This may reflect the fact that, the initial condition evolves differently in the Hall MHD simulation compared to the 
kinetic runs. 

\begin{figure}
\epsfxsize=9cm \centerline{\epsffile{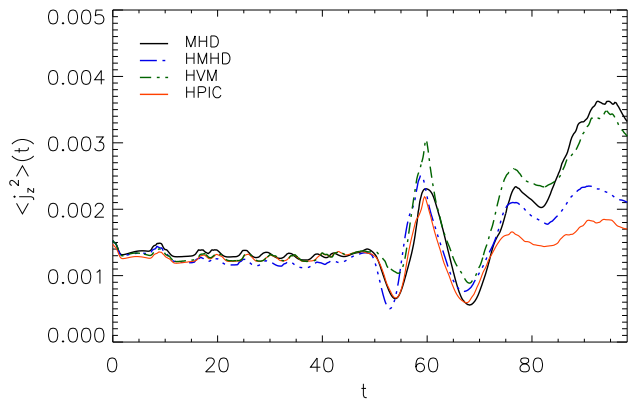}}   
\caption{(Color online) Temporal evolution of $\langle j_z^2\rangle$ for the MHD (black), HMHD (blue), HVM (green) and HPIC 
(red) simulations. For the HPIC simulation, $\langle j_z^2\rangle$ has been smoothed in order to remove particle noise. }
\label{fig:mjz2}
\end{figure}

In order to understand the role of the Hall physics, we computed the normalized generalized cross helicity $\sigma_g=2
e^g/(e^u+e^b)$, where $e^g=0.5\ \langle \mathbf{u}\cdot\mathbf{b}+{\tilde \epsilon} \mathbf{\omega}\cdot\mathbf{u}/2 \rangle $, 
and $\omega={\bf \nabla}\times\mathbf{u}$, which is an invariant of incompressible HMHD \citep{turner86,servidio08}. Figure 
\ref{fig:hchg}(c) displays the temporal evolution of $\sigma_g(t)$ for the MHD (black), the HMHD (dashed blue), HVM (dashed 
green) and HPIC (red) simulations. Note that the evolution of $\sigma_g$ is trivial for the MHD simulation where, 
since $\tilde{\epsilon}=0$, $\sigma_g=\sigma_c$. Moreover, it can be easily appreciated that, for the HMHD case, $\sigma_g$ is 
almost preserved and does not exhibit any significant variation due to the collision itself, even though it shows a slight 
increase in the initial stages of the simulation followed by a decay towards $\sigma_g=0$ [similar to the growth of $\sigma_c$ 
recovered in Fig. \ref{fig:hchg}(b)]. On the other hand the two kinetic cases, which exhibit a similar behavior, show a fast 
growth of $\sigma_g$ in the initial stage of the simulations followed by a decay phase [similar to the growth of $\sigma_c$ 
recovered in Fig. \ref{fig:hchg}(b)]; then, during the collision, $\sigma_g$ significantly increases. We may explain the 
evolution of $\sigma_c$ and $\sigma_g$ as follows. In the MHD run, compressive effects contained in the initial condition as well 
as compressible activity generated during the evolution are not strong enough to break the invariance of $\sigma_c$ (i.e. of 
$\sigma_g$). Instead, in the Hall MHD simulation, the first break of the $\sigma_c$ invariance observed in the initial stage of 
the simulation cannot be associated with the Hall effect since also $\sigma_g$ is not preserved in this temporal region and 
$\sigma_c$ and $\sigma_g$ have a similar evolution. On the other hand, the jump recovered in $\sigma_c$ around $t\simeq\tau=58.9$ 
is significantly related to the Hall physics. In fact, since $\sigma_g$ does not exhibit a similar jump at $t\simeq\tau$, we 
argue that the physics which produces the growth of $\sigma_c$ is the Hall physics (which is taken into account in the invariance 
of $\sigma_g$). Finally, the production of both $\sigma_c$ and $\sigma_g$ recovered in the kinetic simulations cannot be 
completely associated with the Hall effect (which, of course, is still present) but kinetic and compressive effects may have an 
important role.

In order to explore the role of small scales into the dynamics of colliding wave packets, we computed the averaged mean squared
current density $\langle j_z^2\rangle $ as a function of time. This quantity indicates the presence of small scale activity (such 
as production of small scale current sheets), and is reported in Figure \ref{fig:mjz2} for all the simulations. As in the previous 
figures, black, blue dashed, green dashed and red lines refer to the MHD, HMHD, HVM and HPIC cases, respectively. All models show 
a peak of $\langle j_z^2\rangle (t)$ around the collision time $t\simeq \tau$ due to the collision of wave packets. After the 
collision, some high-intensity current activity persists in all the simulations, which present a qualitatively similar evolution 
of $\langle j_z^2\rangle (t)$. 

\begin{figure*}
\epsfxsize=12cm \centerline{\epsffile{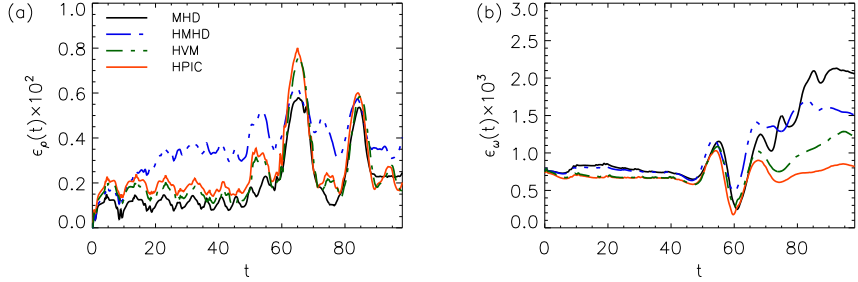}}   
\caption{(Color online) Temporal evolution of $\epsilon_\rho(t)$ (a) and $\epsilon_\omega(t)$ (b). In each panel black, blue, 
green and red lines indicate the MHD, HMHD, HVM and HPIC simulations, respectively. For the HPIC simulation, $\epsilon_\omega(t)$ 
has been smoothed in order to remove particle noise.}
\label{fig:enst}
\end{figure*}

Other quantities that provide physical details about our simulations are $\epsilon_\rho=\langle \delta \rho^2\rangle$ 
(compressibility) and the enstrophy $\epsilon_\omega=\langle \mathbf{\omega}^2\rangle /2$ (fluid vorticity $\omega$). Note that 
$\delta \rho=\rho-\langle  \rho\rangle $. Figure \ref{fig:enst} reports the temporal evolution of $\epsilon_\rho$ (a) and 
$\epsilon_\omega$ (b) for all the runs. Black, blue dashed, green dashed and red lines indicate respectively the MHD, HMHD, HVM 
and HPIC cases. The $\epsilon_\rho$ evolution shows that density fluctuations peak around $t\simeq 63.8$ and $t\simeq 83.4$. The 
two peaks are respectively associated with the collision between the packets and with the propagation of magneto-sonic 
fluctuations generated by the initial strong collision with provide a sort of ``echo'' of the original interaction. Moreover, 
from the initial stage of the simulations, $\epsilon_\rho$ exhibits some modulations, which are produced by the absence of a 
pressure balance in the initial condition. In fact, as packets start to evolve, low-amplitude fast perturbations (clearly visible 
in the density contour plots, not shown here) propagate across the box and collide faster compared to the ``main'' wave packets 
themselves. Moreover, by comparing the different simulations, one notices that, for $t\lesssim 20$, kinetic and Hall runs tend to 
produce a similar evolution of $\epsilon_\rho$, slightly bigger compared to the MHD case. Then, around $t\simeq 20$, the HMHD run 
displays a stronger compressibility with respect to the kinetic cases. This difference is probably due to the presence of kinetic 
damping phenomena which occur in the kinetic cases.

\begin{figure*}
\epsfxsize=\textwidth \centerline{\epsffile{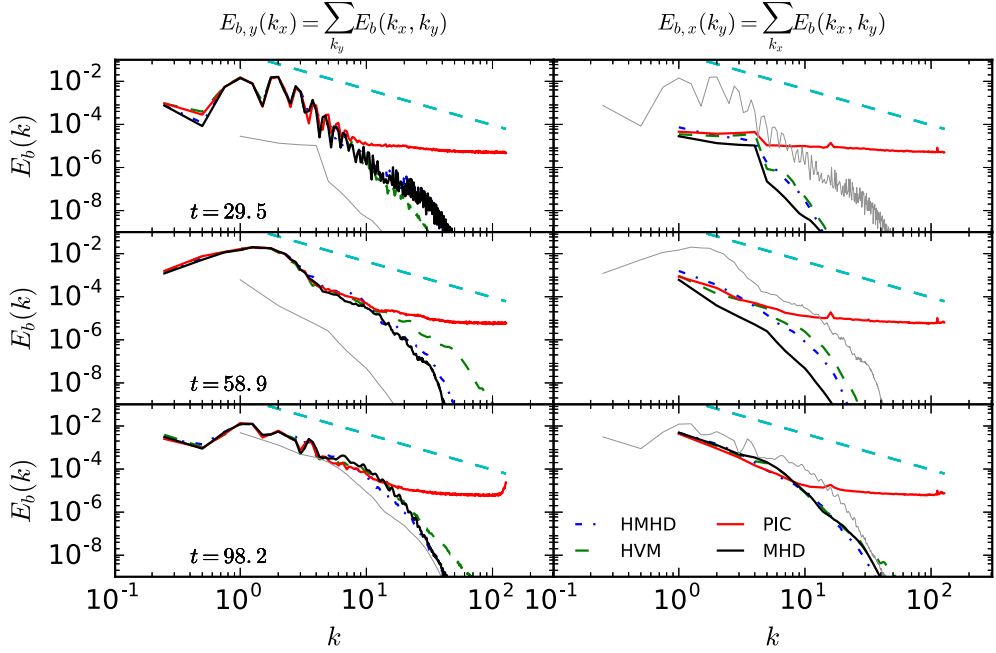}}   
\caption{(Color online) Magnetic energy PSDs $E_{b,y} (k_x) = \sum_{k_y} E_b(k_x,k_y)$ (left column) and $E_{b,x} (k_y) = 
\sum_{k_x} E_b(k_x,k_y)$ (right column) at three time instants: $t=29.5$ (top), $t=\tau=58.9$ (middle) and $t=98.2$ (bottom). In 
each panel black, blue, green and red lines refer to the MHD, HMHD, HVM and HPIC simulations, respectively; while cyan lines 
show the $-5/3$ slope for reference. Moreover, to compare $E_{b,y} (k_x)$ and $E_{b,x} (k_y)$, the gray lines in each panel refer 
only to the MHD simulation and report $E_{b,x} (k_y)$ in the left column and $E_{b,y} (k_x)$ in the right column. }
\label{fig:spectra}
\end{figure*}

The enstrophy $\epsilon_\omega$ is displayed in Fig. \ref{fig:enst}(b). All the runs exhibit a similar evolution of 
$\epsilon_\omega$ up to the wave packet collisions. Then MHD and HMHD cases exhibit a quite similar level of $\epsilon_\omega$, 
slightly bigger compared to the one recovered in the HVM and HPIC cases, where probably kinetic damping does not allow the 
formation of strong vortical structures at small scales.

\begin{figure}
\epsfxsize=10cm \centerline{\epsffile{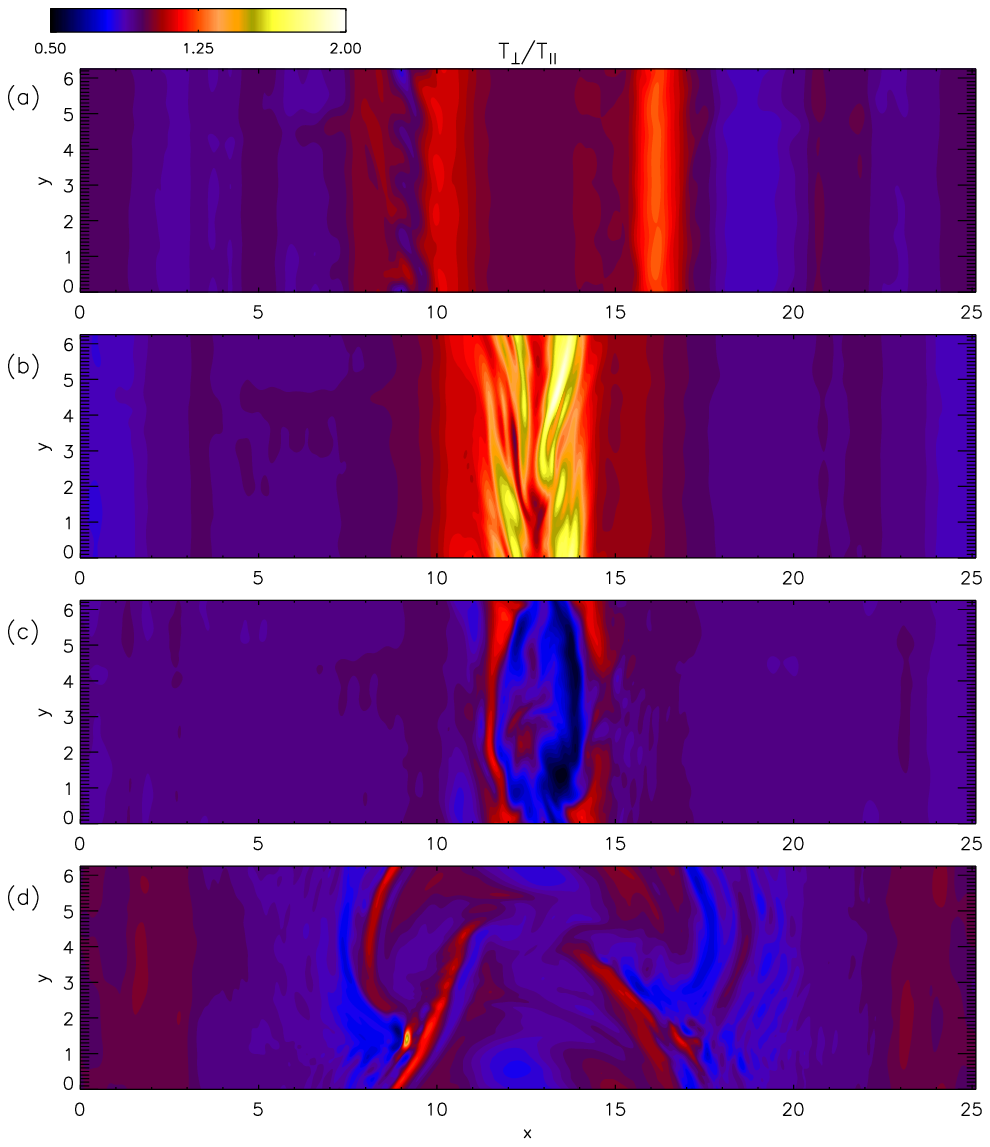}}   
\caption{(Color online) Contour plots of the temperature anisotropy, for the HVM run, evaluated in the LBF at four time instants: 
(a) $t=29.5$, (b) $t=\tau=58.9$, (c) $t=70.7$ and (d) $t=98.2$. }
\label{fig:anis}
\end{figure}

It is interesting to compare different simulations also by looking at power spectral densities (PSDs). Figures \ref{fig:spectra} 
show the magnetic energy PSD integrated along $k_y$ $E_{b,y} (k_x) = \sum_{k_y} E_b(k_x,k_y)$ (left column) and along $k_x$ 
$E_{b,x} (k_y) = \sum_{k_x} E_b(k_x,k_y)$ (right column); while each row respectively refers to  $t= 29.5$ (top row), 
$t=\tau=58.9$ (center row) and $t=98.2$ (bottom row). The cyan dashed line shows the $k^{-5/3}$ slope for reference while, in 
each panel, black, blue, dashed green and red lines indicate respectively MHD, HMHD, HVM and HPIC simulations. Moreover, to 
compare the two wave-number directions, gray lines in each panel report the corresponding PSD obtained from the MHD run, reduced 
in the other direction [for example, in the top row left panel, the gray line refers to $E_{b,x} (k_y)$ for the MHD simulation 
while other curves in th same panel report $E_{b,y} (k_x)$]. It is interesting to note that, at $t=29.5$, all the simulations 
exhibit a steep spectrum in $E_{b,y}(k_x)$, related to the initial condition, while the difference in power between $E_{b,y}(k_x)$ 
and $E_{b,x}(k_y)$ - the latter being significantly smaller than the former - tends to reduce in the final stages of the 
simulations. This suggest, as described in Paper I, the presence of nonlinear couplings which cause spectra to become more isotropic. 
Moreover the comparison between the different simulations indicates that the dynamics at large scales is described in a quite 
similar way by all runs, while  at small scales some differences are revealed. In particular, the HPIC simulation is affected by 
particles noise while the HVM run contains more energy at small scales compared to MHD and HMHD. Note that the presence, 
in MHD and HMHD runs, of an explicit resistivity prevents the population of small scales.

\begin{figure}
\epsfxsize=10cm \centerline{\epsffile{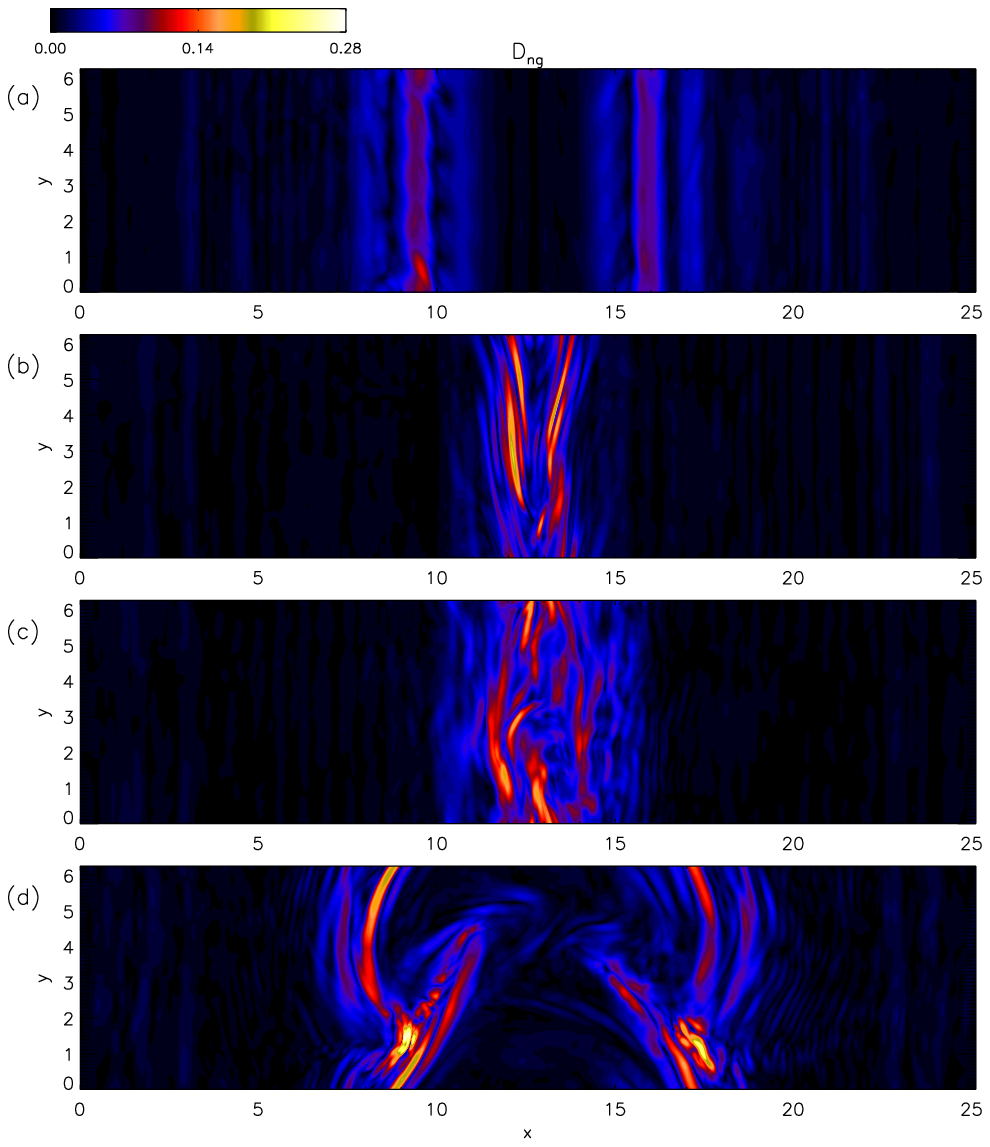}}   
\caption{(Color online) Contour plots of the degree of temperature non-gyrotropy $D_{ng}$, for the HVM run, evaluated in the LBF 
at four time instants:  (a) $t=29.5$, (b) $t=\tau=58.9$, (c) $t=70.7$ and (d) $t=98.2$.}
\label{fig:Dng}
\end{figure}

To summarize this section, we compared our numerical codes by analyzing some global diagnostics and we conclude that the 
Moffatt-Parker scenario is quite well satisfied by MHD. However, other intriguing characteristics are observed when one moves 
beyond the MHD treatment. Moreover, the comparison between kinetic codes suggests that HVM and HPIC simulations display 
qualitatively similar features at large scales. However, when one aims to analyze the dynamics at small scales, HPIC simulations 
suffers from thermal particle noise. Magnetic energy spectra differ in the HPIC case compared with the HVM case. Moreover, by 
comparing the $j_z$ contour plots, one can easily appreciate how HPIC simulation is affected by particle noise. Based on these 
considerations, we continue the analysis of the kinetic features produced in Alfv\'en wave packet collisions by focusing 
only on the HVM case. 

\section{Kinetic features recovered during the wave packets interaction}
\label{sec:kin}

We begin a description of the kinetic signatures present in the Vlasov simulation by looking at the temperature anisotropy. 
Fig. \ref{fig:anis} reports the contour plots of the temperature anisotropy $T_\perp/T_\parallel$, where the parallel and 
perpendicular directions are evaluated in the local magnetic field frame (LBF), at four time instants: $t=29.5$ (a), 
$t=\tau=58.9$ (b), $t=70.7$ (c) and $t=98.2$ (d). Clearly temperature anisotropy is present even before the main wave packets 
collision, due to the fact that the initial wave packets are not linear eigenmodes of the Vlasov equation and, hence, 
their dynamical evolution leads to an anisotropy production. Moreover, a more careful analysis suggests that the left wave packet 
tends to produce regions where $T_\perp/T_\parallel < 1$ close to the packet itself (which, as it can be appreciated in Fig. 
\ref{fig:jz}, is localized around $x\simeq9.5$), while the right wave packet (localized around $x=15.7$) is characterized by 
$T_\perp/T_\parallel > 1$. The presence of different temperature anisotropies is related to the broken symmetry with respect 
to the center of the $x$ direction. Indeed, the dynamics of the wave packets is different if they move parallel or anti-parallel 
to $B_{0,x}$. This produces the different temperature anisotropy recovered in the top panel of Fig. \ref{fig:anis}.

When the packets collide [Fig. \ref{fig:anis} (b)], sheets characterized by a strong temperature anisotropy ($T_\perp/T_\parallel 
> 1$) are recovered, correlated spatially with the current density structures. Then, at $t=70.7$ [Fig. \ref{fig:anis} (c)], wave 
packets split again and a region, localized at $(x,y) \simeq (14.3,1.0)$, where the temperature anisotropy suddenly moves from 
values $T_\perp/T_\parallel < 1$ towards $T_\perp/T_\parallel > 1$ ones is present. We will show that this region also shows the 
presence of strong departures from the equilibrium Maxwellian shape. At the final stage of the simulations [Fig. \ref{fig:anis} 
(d)], each wave packet continues traveling, accompanied by a persistent level of temperature anisotropy, which is, indeed, well 
correlated with the current structures.

\begin{figure}                         
\epsfxsize=7cm \centerline{\epsffile{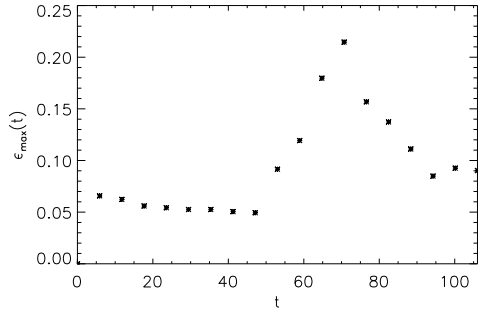}}   
\caption{(Color online) Temporal evolution of $\epsilon_{max} (t)$ for the HVM run. }
\label{fig:maxeps}
\end{figure}

It is interesting to point out that, beyond the presence of temperature anisotropies, regions characterized by a temperature 
nongyrotropy are also recovered. Many methods have been proposed by evaluating the temperature nongyrotropy 
\citep{aunai13,swisdak16}. Here we adopt the measure $D_{ng}$ \citep{aunai13}, which is proportional to the root-mean-square of 
the off-diagonal elements of the pressure tensor. Fig. \ref{fig:Dng} shows the contour plots of $D_{ng}$ at four time instants: 
$t=29.5$ (a), $t=\tau=58.9$ (b), $t=70.7$ (c) and $t=98.2$ (d). Moreover, as for the temperature anisotropy, the evolution of the 
two wave packets tends to produce temperature nongyrotropies also before the wave packets collision [Fig. \ref{fig:Dng}(a)]. Then, 
during the collision [Fig. \ref{fig:Dng}(b)--(c)], the temperature nongyrotropy becomes more intense and it is also quite well 
correlated with the current structures. At the final stage of the simulation [Fig. \ref{fig:Dng}(d)], each wave packet is connoted 
by a level of nongyrotropy which is quite bigger compared to the value before the collision. The presence of nongyrotropic 
regions suggests that it is fundamental to retain a full velocity space where the VDF is let free to evolve and, eventually, 
distort. 

To further support the idea that kinetic effects are produced during the interaction of the wave packets, we computed the $L^2$ 
norm difference \citep{servidio12,greco12,servidio15}:
\begin{equation}
 \epsilon(x,y,t) = \frac{1}{n} \sqrt{\int \left[ f({\bf x}, {\bf v}, t) - f_M({\bf x}, {\bf v}, t)\right]^2 d{\bf v} }
\end{equation}
which measures the displacements of the proton VDF $f({\bf x}, {\bf v}, t)$ with respect to the associated Maxwellian distribution 
function $f_M({\bf x}, {\bf v}, t)$, built such that density, bulk speed and total temperature of the two VDFs are the same. 
Figure \ref{fig:maxeps} reports the evolution of the $\epsilon_{max}(t)=max_{(x,y)} \epsilon(x,y,t)$ as a function of time. 
Clearly, as for the previous proxies of kinetic effects, also $\epsilon_{max}$ moves away from zero in the early phases of the 
simulation due to the fact that the initial-condition is not a Vlasov eigenmode. Moreover, after the first initial variation, 
$\epsilon_{max}$ is almost constant until wave packets interact. Then, during the collision, $\epsilon_{max}$ grows and reaches 
its maximum at $t=70.7$, quite later than the wave packet collisions. Then it decreases and saturates at a value about two times 
bigger than the value before the collision, thus suggesting, again, that there is ``net'' production of non-Maxwellian features in 
the VDF during the wave packets interaction.

\begin{figure*}               
 \centering
 \begin{minipage}{0.65 \textwidth}
  \epsfxsize=9cm \centerline{\epsffile{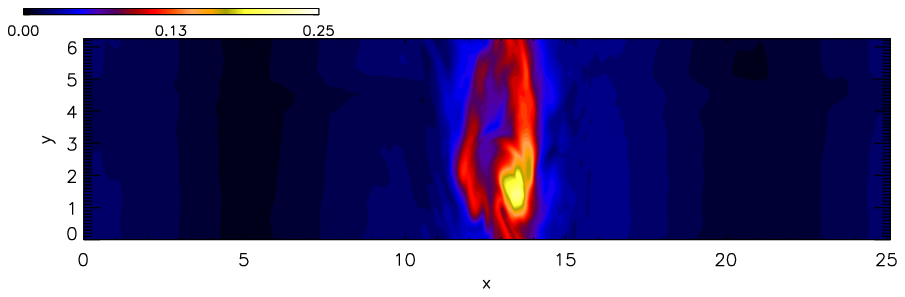}}   
 \end{minipage}
 \hfill
 \begin{minipage}{0.33 \textwidth}
  \epsfxsize=4cm \centerline{\epsffile{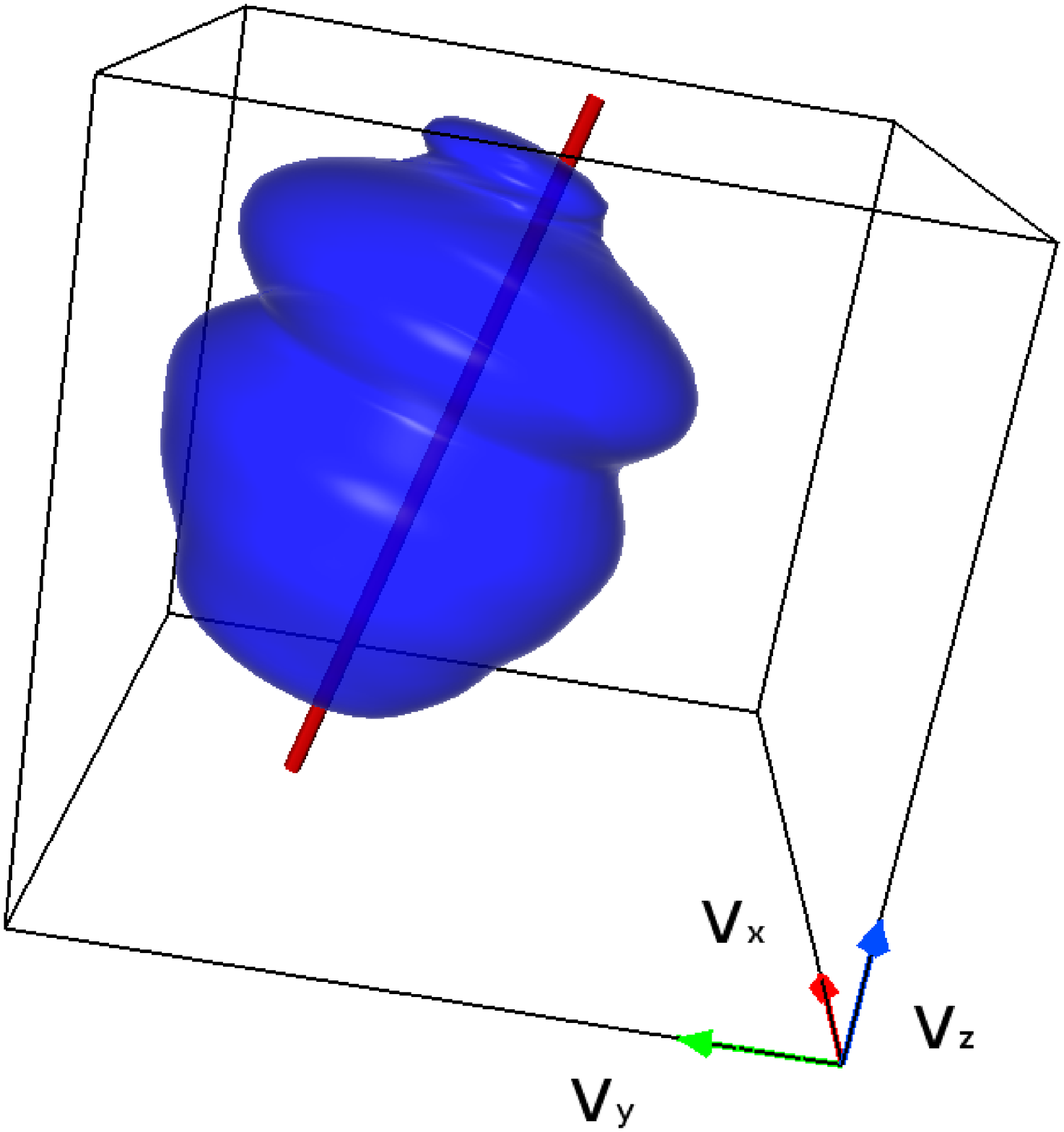}}   
 \end{minipage}
 \caption{(Color online) (Left) Contour plots of $\epsilon (t)$, for the HVM run, at $t=70.7$. (Right) Proton distribution 
functions, in the spatial point $(x^*,y^*)$ where $\epsilon(x^*,y^*,t)=\max_{(x,y)} \epsilon(x,y,t)$ at $t=70.7$. The local 
magnetic  field direction is indicated by a red line. }
 \label{fig:epsVDF}
\end{figure*}

In order to appreciate the structure of $\epsilon$ in the spatial domain, the left panel of Figure \ref{fig:epsVDF} shows the 
contour plot of $\epsilon(x,y,t)$ at the time instant $t=70.7$ (when $\epsilon$ reaches its maximum value). The $\epsilon$ 
contours are correlated with the current structures and with the anisotropic/nongyrotropic regions. Moreover, a blob-like region 
where $\epsilon$ reaches its maximum is present. This area is associated with the region where the temperature anisotropy moves 
from $T_\perp/T_\parallel < 1$ values towards $T_\perp/T_\parallel > 1$ ones [See Fig. \ref{fig:anis}(c)]. In this area the VDF 
strongly departs from the Maxwellian. The right panel of Fig. \ref{fig:epsVDF} shows the three dimensional isosurface plot of the 
VDF at $t=70.7$ and in the spatial point $(x^*,y^*)$ where $\epsilon(x^*,y^*,t=70.7)=\max_{(x,y)} \epsilon(x,y,t=70.7)$. A strong 
beam, parallel to the local magnetic field direction, is observed in the VDF in Fig. \ref{fig:epsVDF}. The drift speed of the beam 
is about ${\tilde c}_A$. The production of a beam due to the interaction of two wave packets has been also pointed out by 
\citet{he15}.

\section{Conclusion}
\label{sec:concl}

In this paper we have described the interaction of two Alfv\'enic wave packets by means of MHD, Hall MHD and hybrid kinetic 
simulations of the same physical configuration. Kinetic simulations have been performed with two different codes: an Eulerian 
Vlasov-Maxwell code \citep{valentini07} and hybrid Particle-in-cell code \citep{ParasharPP09}. By approaching the {\it Parker} \& 
{\it Moffatt} problem within several physical frameworks, we have comparatively analyzed different effects (compressive, Hall and 
kinetic) which contribute to the general, complex puzzle.

The analysis performed in Paper I where MHD and Vlasov simulations were compared has been here extended by including the Hall 
MHD framework. In particular, we showed how, moving beyond the pure MHD treatment, dispersion as well as kinetic effects play an 
important role. Furthermore, the analysis of HMHD and kinetic runs allows to separate the presence of dispersive and purely 
kinetic features. It is also interesting to note that the compressive activity is different in the Hall case compared to the 
kinetic runs, indicating that some kinetic damping processes may be active in the Vlasov simulation. A separate type of comparison 
is afforded by comparative analysis of the HPIC and HVM runs, While these methods should describe, approximately, the same 
physics, i.e., the Vlasov treatment of collsionless plasma dynamics, the comparison between the different codes is interesting 
from a methodological perspective, and therefore represents a contribution to the Turbulent Dissipation Challenge 
\citep{ParasharJPP15}. In particular, the two kinetic simulations performed are able to take into account the dynamics which 
occurs at large spatial scales and their comparison is successful in this range of scales. However the PIC runs lacks accuracy 
when smaller spatial scales are produced by the collision of the two wave packets, thus indicating that the Eulerian approach 
better describes the dynamics of the system at small spatial scales. Of course the comparison is expected to become better if the 
number of particle per cell in the PIC simulation gets bigger \citep{camporeale11,franci15}.

Based on the last consideration, we have analyzed the production of kinetic signatures by focusing only on the HVM simulation. 
Several proxies which are routinely adopted for highlighting the presence of kinetic features indicate that wave packets 
tend to produce kinetic effects such as temperature anisotropies and nongyrotropies also before the main wave packets 
interaction. This is related to the fact that the initial condition, consisting of quasi-Alfv\'enic wave packets, is neither a 
Vlasov equilibrium nor a Vlasov eigenmode. Therefore it dynamically leads to the production of kinetic features. 

The analysis of kinetic effects before and after the main wave packets collision indicates that some kinetic features are enhanced 
by the collision itself and each wave packet is significantly characterized by a strong degree of non-thermal signatures. In 
particular the presence of temperature nongyrotropies suggests that descriptions based on reduced velocity space assumptions may 
partially fail the description of such features. Finally, similarly to the observations of \citet{he15}, during the wave packet 
collision, a beam in the velocity distribution function is observed to form along the direction of the local magnetic field. This 
characteristic may connect our results with the general scenario of wave packets observed in natural plasmas such as the solar 
wind.

Research is supported by NSF AGS-1063439, AGS-1156094 (SHINE), AGS-1460130 (SHINE), and NASA grants NNX14AI63G (Heliophysics 
Grandchallenge Theory), and the Solar Probe Plus science team (ISIS/SWRI subcontract No. D99031L), and Agenzia Spaziale Italiana 
under the contract ASI-INAF 2015-039-R.O “Missione M4 di ESA: Partecipazione Italiana alla fase di assessment della missione 
THOR”. Kinetic numerical simulations here discussed have been run on the Fermi supercomputer at Cineca (Bologna, Italy), within 
the ISCRA-C project IsC37-COLALFWP (HP10CWCE72) and on the Newton parallel machine at University of Calabria (Rende, Italy).

\bibliographystyle{jpp}

\end{document}